\documentclass[aps,prb,10pt,twocolumn,raggedbottom,nobalancelastpage]{revtex4-1}
\usepackage{amsmath}
\usepackage{graphicx}

\newcommand{\dvec}[1]{\ensuremath{\boldsymbol{#1}}}
\newcommand{\ve}{\dvec{\mathrm{e}}}
\newcommand{\vj}{\dvec{\mathrm{j}}}
\newcommand{\vell}{\dvec{\ell}}
\newcommand{\vp}{\dvec{\mathrm{p}}}
\newcommand{\vs}{\dvec{\mathrm{s}}}
\newcommand{\vS}{\dvec{\mathrm{S}}}
\newcommand{\vsigma}{\dvec{\sigma}}

\begin{document}
\title{Optical manifestations of symmetry breaking in bilayer graphene}
\author{D. S. L. Abergel}
\affiliation{Condensed Matter Theory Center, Department of Physics,
University of Maryland, College Park, MD 20742, USA}
\author{Vladimir I. Fal'ko}
\affiliation{Department of Physics, Lancaster University, Lancaster, LA1
4YB, United Kingdom}

\begin{abstract}
	We propose a spectroscopic method of identifying broken symmetry
	states of bilayer graphene.
	We demonstrate theoretically that, in contrast to gapped states, a
	strained bilayer crystal or nematic phase of the electronic liquid
	are distinguishable by the dependence of the lineshape of absorption
	on the polarization of the light. 
	This property is characteristic for both the infrared and
	far-infrared spectral ranges, which correspond to the absorption by
	transitions between low-energy bands and split bands, and
	transitions between the low-energy valence and conduction bands,
	respectively.
\end{abstract}

\maketitle

In this Rapid Communication we study how symmetry breaking in bilayer
graphene (BLG) \cite{mccann-prl96} 
manifests itself in the optical spectra in the infrared (IR) and
far-infrared (FIR) spectral ranges.\cite{abergel-prb75, nicol-prb77,
wang-sci320, ohta-sci313, zhang-nature459, mak-prl102, kuzmenko-prb80,
li-z-prl102}
Symmetry breaking in BLG may be caused both by external perturbations and
by internally developed instabilities generated by the electron--electron
interactions. 
For example, strain in BLG which might be inflicted on the crystal
involuntarily upon thermal annealing and cooling of suspended BLG
devices would asymmetrically change the topology of the low-energy
dispersion.\cite{muchakruczynski-prb84, pereira-prb80, dietl-prl100}
Also, by applying a perpendicular electric field, one breaks the
inversion symmetry of the lattice opening an externally tunable gap
between the valence and conduction bands.\cite{mccann-prl96}
Alternatively, at low temperatures, undoped pristine BLG may undergo a
spontaneous symmetry breaking transition into one of the recently
discussed strongly correlated ground states.\cite{nilsson-prb73,
zhang-prb81, nandkishore-prb82, nandkishore-prl104, jung-prb83,
zhang-prl106, min-prb77, kharitonov-arXiv1109, vafek-prb82, vafek-prb81,
lemonik-prb82} In particular, the phases favored by the renormalization
of short-range electron--electron interaction constants
\cite{vafek-prb81, lemonik-prb82, lemonik-prb85} are
a nematic state in which the isotropy of the band structure is
reduced in a similar way to strained BLG, \cite{mayorov-sci333,
muchakruczynski-prb84} and gapped layer-antiferromagnetic
\cite{kharitonov-arXiv1109} and spin flux phases.
\cite{lemonik-prb85}
Although several transport experiments \cite{martin-prl105,
mayorov-sci333, velascojr-natnano7, freitag-prl108} reported
observations of some broken symmetry states in BLG, the exact nature of
the ground state still remains to be established.

Here, we show how infrared (IR) and far-infrared (FIR) absorption spectroscopy can be used to distinguish
between some of the broken symmetries.
The feature discussed below is that strain (or a phase transition to the
nematic state) induces a dependence of absorption of light on the
polarization of the radiation. This anisotropy can be characterized by a
factor
\begin{equation}
	Q = \frac{g_\parallel - g_\perp}{g_\parallel + g_\perp},
	\label{eq:Q}
\end{equation}
where $g_\parallel$ ($g_\perp$) is the absorption coefficient of light
with linear polarization $\ve$ parallel (perpendicular) to the principal
strain axis (or the direction chosen by the order parameter of the
nematic phase). 
In contrast, the gapped phases show isotropic absorption and notable 
qualitative differences in the lineshape of their absorption
spectrum as compared to the strained (nematic) and unperturbed BLG states.

The absorption coefficient \cite{abergel-prb75, nicol-prb77} analyzed
in this study,
\begin{equation}
	g_{\ve}(\omega) = \frac{4\pi\hbar g_s}{c \omega \mathcal{A}} 
	\mathrm{Im}
	\sum_{\substack{\vp,\lambda,\lambda'\\\alpha,\beta}}
	\frac{f(\vp\lambda') - f(\vp\lambda)}
	{\omega + \epsilon_{\vp\lambda} - \epsilon_{\vp\lambda'} + i0} 
	e_\alpha M_{\alpha\beta}^{\lambda\lambda'}
	e^\ast_\beta,
	\label{eq:absorp}
\end{equation}
is given by the ratio of the Joule heating and the flux of incident
radiation.
Equation \eqref{eq:absorp} describes transitions between initial and
final plane wave states marked by indices $\lambda$ which include
the band, branch, and valley (spin is also taken into account);
$\epsilon_{\vp\lambda}$ is the energy of a plane wave with momentum
$\vp$, $f$ is the Fermi function, $\mathcal{A}$ is the normalization
area, and
\begin{equation}
    M_{\alpha\beta}^{\lambda\lambda'} = \langle \vp\lambda |
	\hat{j}_\alpha^\dagger | \vp\lambda'\rangle
	\langle \vp\lambda' | \hat{j}_\beta | \vp\lambda \rangle
	\label{eq:M}
\end{equation}
are determined by the current density operator
$\hat{\vj}=e\partial_{\vp}\mathcal{H}$.
Since the momentum transferred by light is negligibly small, in
$M_{\alpha\beta}^{\lambda\lambda'}$ the momenta of electrons in the
initial and final states of the inter-band transitions are taken to be
equal.
The corresponding plane wave state wave functions are the four-component
eigenstates $\left| \ldots \right\rangle \propto
\left( \psi_{A_1}, \xi\psi_{B_2}, \psi_{A_2}, \xi \psi_{B_1} \right)$
of the Hamiltonian \cite{mccann-prl96,
muchakruczynski-prb84}
\begin{equation}
	\mathcal{H} = \begin{pmatrix} 0 & v \vsigma\cdot\vp \\
	v \vsigma\cdot\vp & \gamma_1 \sigma_x \tau_z \end{pmatrix} + \delta
	H,
	\label{eq:Ham}
\end{equation}
where $A_{1(2)}$ and $B_{1(2)}$ identify the sublattices of the
honeycomb lattices in the top (bottom) layers, $\xi=\pm$ distinguishes
the $K$ and $K'$ corners of the hexagonal Brillouin zone, and Pauli
matrices $\sigma_x$, $\sigma_y$, $\sigma_z$, and $\tau_z$ act on the
sublattice and valley components of $|\ldots \rangle$, respectively.
Also, $\gamma_1$ stands for the inter-layer coupling between atoms on
$A_2$ and $B_1$ lattice sites, and $v$ is the Dirac velocity for
monolayer graphene. 

For unperturbed BLG ($\delta H=0$), Eq.~\eqref{eq:Ham} determines
\cite{mccann-prl96} a pair of low-energy bands near the Brillouin zone
corners with spectrum
$\epsilon_{\vp\pm} = \pm p^2/2m^\ast \equiv \pm \epsilon_0$,
$m^\ast = \gamma_1/2v^2 \approx 0.035 m_e$ for states located mainly on
sublattices $A_1$ and $B_2$.
Equation \eqref{eq:Ham} also determines two split bands with quadratic
dispersion $\epsilon_{\vp\pm} \approx \pm(\gamma_1 + \epsilon_0)$ and
wave functions that have equal weight on lattices $A_2$ and $B_1$. 
Note that the split band states are almost unperturbed by the
strain or the formation of a nematic phase or one of the gapped phases.
\cite{mccann-prl96}

For the sake of convenience, the ``valley momentum'' $\vp$ in
Eqs.~\eqref{eq:M} and \eqref{eq:Ham} is determined with respect to the
position of the Dirac point in the
graphene monolayer, which is shifted from the Brillouin zone corners $K$
and $K'$ in a homogeneously strained crystal.\cite{guinea-natphys6} In
monolayer graphene, such a shift $\vp\to \vp' + \dvec{\mathrm{a}} \equiv
\vp$ in the momentum space is trivially absorbed into a gauge
transformation so that it does not influence observable
characteristics, such as the absorption spectrum. 
In contrast, in BLG the interplay of such a shift and the interlayer
skew hopping $\gamma_3$ which couples sublattices $A_1$ and $B_2$
generates a perturbation which cannot be eliminated from the Hamiltonian
by any gauge transformation. 
This, as well as other possible symmetry-breaking perturbations in BLG
are included in
\begin{equation*}
	\delta H = \begin{cases}
	w \tau_z \vsigma \cdot \dvec{\Lambda}, &
		\text{strain/nematic,
		\cite{muchakruczynski-prb84, vafek-prb81,
		lemonik-prb82, lemonik-prb85}} \\
	u \tau_z \sigma_z, & \text{layer asymmetry,
		\cite{mccann-prl96, nandkishore-prl104, jung-prb83}} \\
	u (\vs\cdot\vS) \tau_z \sigma_z, &
	\text{antiferromagnetic, \cite{lemonik-prb85,
	kharitonov-arXiv1109, min-prb77, vafek-prb82}} \\
	u (\vs\cdot\vS) \sigma_z, & \text{spin flux.
	\cite{lemonik-prb85}}
	\end{cases}
\end{equation*}

The first term in $\delta H$ accounts for the effect of strain
\footnote{For strain \cite{muchakruczynski-prb84}
$w=\gamma_3(\delta-\delta')\left( \frac{d \ln \gamma_3}{dR_{AB}} -
\frac{d \ln \gamma_0}{dR_{AB}} \right)$ where $\delta$ and
$\delta'$ are eigenvalues of the strain tensor, and $R_{AB}$ is the
closest neighbor distance between carbon atoms in the monolayer.}
or nematic order, with the direction of the unit vector
$\dvec{\Lambda} = (\cos 2\theta, \sin 2\theta)$ set by the direction
$\vell = (\cos\theta, \sin\theta)$ of the principal stretching
direction.
This perturbation changes the low-energy electron dispersion into
\begin{equation}
	\epsilon_{\vp\pm} = \pm \frac{1}{2m} |\vp - \sqrt{2mw}\vell| \times
	|\vp + \sqrt{2mw} \vell|,
\end{equation}
which features a Lifshitz transition at $\epsilon = \pm |w|$, from an
almost parabolic dispersion at $|\epsilon|\gg |w|$ to a pair of Dirac
cones at $\vp = \pm \sqrt{2mw}\vell$ shifted from each other along the
anisotropy axis $\vell$.
In the following, the axis $\vell$ will be used as the reference
direction to distinguish between ``parallel'', $\ve \parallel \vell$, and
``perpendicular'', $\ve \perp \vell$, polarizations of light and the
corresponding absorption coefficients $g_\parallel$ and $g_\perp$ in
Eq.~\eqref{eq:Q}. 

Also, in $\delta H$, antiferromagnetic (AF) and spin-flux (SF) states
are characterized by a splitting $2u$ between the valence
and conduction bands at the $K$ point and have identical spectra
$\epsilon_{\vp\pm}=\pm\sqrt{u^2 + \epsilon_0^2}$, the same as in graphene
with an asymmetry gap opened, for example, by a perpendicular electric
field.  \cite{mccann-prl96}
For magnetic phases of BLG, $\vS$ characterizes the spin quantization
axis, but  since the current operator $\hat{\vj}$ does not depend on the
spin at all, these BLG states will behave identically in the optical
absorption.

\begin{figure}
	\includegraphics[]{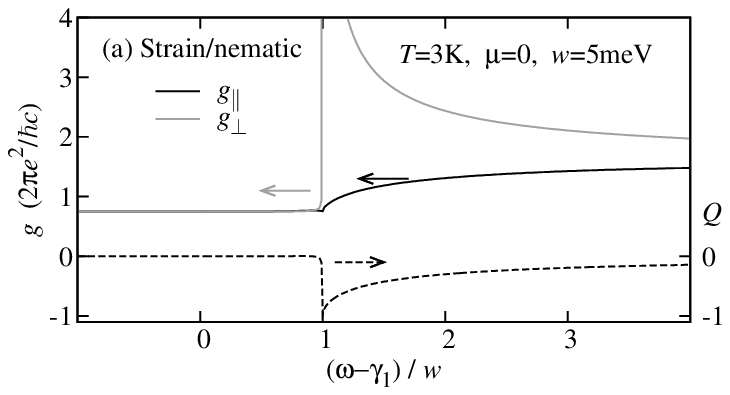}
	\includegraphics[]{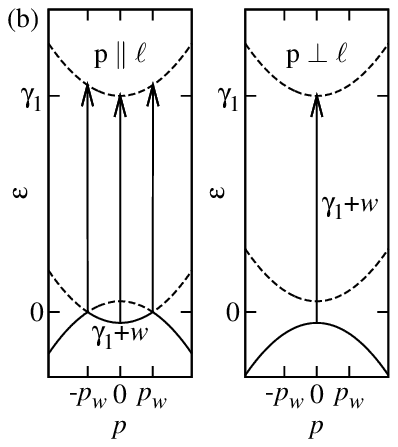}
	\includegraphics[]{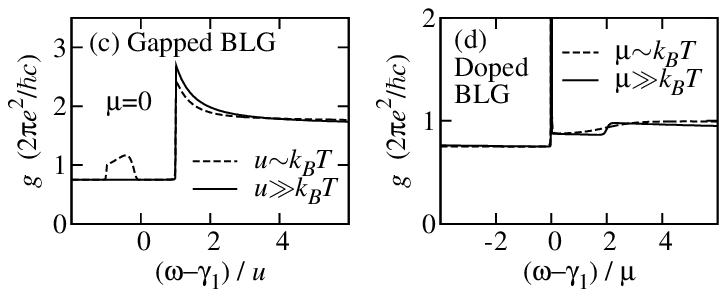}
	\includegraphics[]{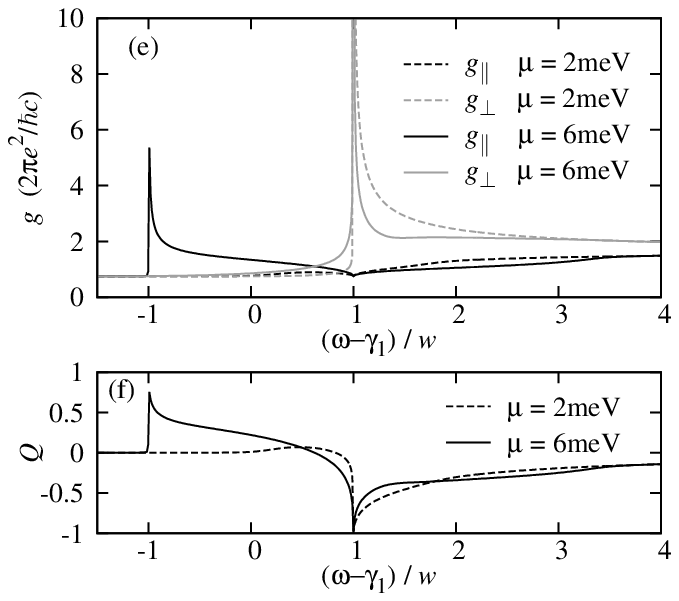}
	\caption{(a) Absorption coefficients $g_\parallel$ and $g_\perp$ and
	absorption anisotropy $Q$ for 
	undoped strained/nematic bilayer graphene with $w=5\mathrm{meV}$ and
	$T=3\mathrm{K}$. 
	(b) Sketch illustrating optical transitions from the low-energy
	bands at $\mu=0$ and $T=0$ in strained BLG or the nematic phase of
	BLG. 
	Solid (dashed) lines denote filled (empty) states. The two extremal
	parts of the dispersion are shown for $\vp \parallel \vell$ (with
	Dirac points at $\pm p_w$ with $p_w = \sqrt{2mw}$), and $\vp
	\perp \vell$.  The arrows mark the range of $p$ for which the
	threshold transitions with $\omega=\gamma_1+w$ exist.
	(c) The absorption coefficient for undoped, gapped bilayer graphene.
	(d) The absorption coefficient for the unperturbed system with
	finite chemical potential $\mu$.
	(e) Absorption coefficient and (f) anisotropy $Q$ of
	strained BLG with finite doping for $w=5\mathrm{meV}$ and
	$T=3\mathrm{K}$, for $\mu=2\mathrm{meV}<w$ and $\mu=6\mathrm{meV}>w$.
	\label{fig:g1en}}
\end{figure}

In the IR spectral range, $\omega \approx \gamma_1 \approx
0.4\mathrm{eV}$, where $\omega$ is the energy of the incident light, the
optical transitions which are sensitive to the BLG symmetry breaking are
those between the small momentum parts of the split bands and the
low-energy bands.
In addition, the spectral density of IR absorption includes the
contribution from the transitions between the two low-energy bands at
high momentum which provides an almost-constant background in the
absorption spectrum.\cite{abergel-prb75, nicol-prb77}
In Fig.~\ref{fig:g1en}(a) we show the calculated absorption spectrum 
for the two characteristic polarizations of light and the polarization
factor $Q$ for undoped, strained BLG with $w=5\mathrm{meV}$ at
$T=3\mathrm{K}$. 
The absorption spectrum of BLG in the nematic phase would have the same
features: the anisotropy of the absorption for
$\omega \approx \gamma_1 + w$, where the absorption for $\ve \perp \vell$
(gray line) shows a strong peak near the threshold whereas absorption
for $\ve \parallel \vell$ (black line) is weak:
\begin{equation*}
	g_\perp \propto (\omega-\gamma_1-w)^{-2}, \quad g_\parallel \to 0.
\end{equation*}
This occurs because the matrix elements $M$ in Eq.~\eqref{eq:M}
preferentially select transitions with $\vp \perp \vell$ for
$g_\parallel$ and $\vp \parallel \vell$ for $g_\perp$. 
In the first case, when $\vp \perp \vell$, only transitions from right
at the center of the $K$ point have the threshold energy $\gamma_1+w$, 
but in contrast,
when $\vp \parallel \vell$, there is a finite range of momenta,
$-\sqrt{2mw} < p < \sqrt{2mw}$, where the low-energy valence
band and conduction split band are shifted on the energy scale by
exactly $\gamma_1 + w$, [Fig.~\ref{fig:g1en}(b)].
(This is also true for the valence split band and the low-energy
conduction band).
This produces a singularity at the interband absorption edge,
$\gamma_1+w$.
\footnote{Since this divergence is created by the existence of
parallel bands, it may be cut off by any additional perturbation which
modifies the slope of one band with respect to the other. For example,
the next-nearest neighbor inter-layer hops, the inter-layer hops from
dimerized to un-dimerized lattice sites, or a momentum dependence of
$\gamma_1$ (and therefore of the effective mass) can all cause this
effect.}

For comparison, in Fig.~\ref{fig:g1en}(c) we show the absorption
spectrum characteristic for any of the gapped states of BLG. 
Here, the absorption coefficient does not depend on the polarization of
the photon and there is a feature at the threshold $\omega = \gamma_1+u$
of the lowest energy inter-band transitions. 
The height of this peak is constant for $u\gg
k_B T$. When $k_BT \gtrsim u$ (this situation is considered having in
mind the gapped state caused by the inter-layer asymmetry due to
external perturbation rather than an intrinsic phase transition) thermal
occupation of the low-energy conduction band also allows transitions to
the split band with energy $\gamma_1 - u$, which yields a small
additional polarization-independent peak in $g(\omega)$.
The absorption of IR light by unperturbed BLG with finite doping and
chemical potential $\mu$ is shown in Fig.~\ref{fig:g1en}(d),
\footnote{For doped BLG, there is a step at $\omega = \gamma_1 \pm 2\mu$
because, for $\gamma_1 < \omega < 2\mu$, the transitions between the split
valence band and low-energy conduction band are Pauli-blocked.
There is also a very sharp spike caused by the transitions between the two
parallel conduction bands (low-energy band and split band at $\gamma_1$)
for $k<\sqrt{2m\mu}/\hbar$.} in precise agreement with previous
calculations of the optical conductivity.\cite{abergel-prb75,
nicol-prb77, kuzmenko-prb79}

Figure \ref{fig:g1en}(e) shows the absorption spectrum of
strained BLG with finite doping (the chemical potential $\mu\neq 0$
is counted from the Dirac point energy). Note that the nematic phase is
not expected to survive at finite doping.
When $|\mu|<w$ (dashed line), the absorption spectrum remains almost
unchanged as compared to the undoped case, but for $|\mu|>w$ (solid
line) a new peak appears at $\omega=\gamma_1-w$ in the $\ve \parallel
\vell$ polarization but not in the $\ve \perp \vell$ polarization.
This occurs because, with this level of doping, transitions from the
low-energy conduction band above the Lifshitz transition become
accessible.
This asymmetry manifests itself in the polarization degree $Q$ of the
absorption, shown in Fig.~\ref{fig:g1en}(f) for $|\mu|<w$
(dashed line) and $|\mu|>w$ (solid line).

\begin{figure}
	\includegraphics[]{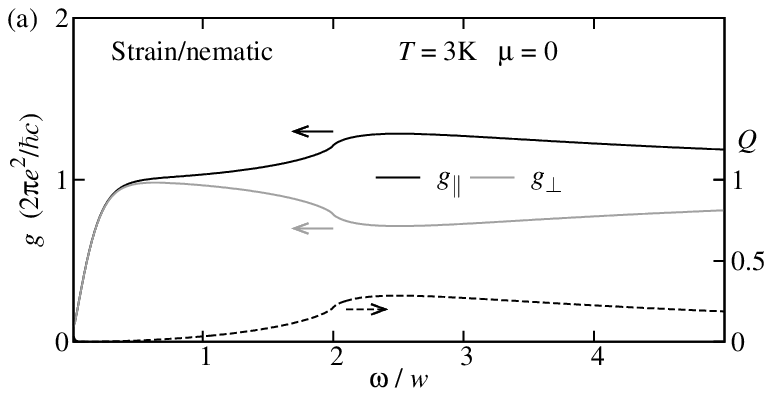}
	\includegraphics[]{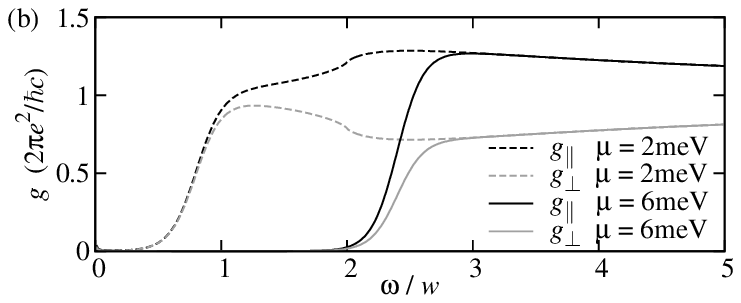}
	\includegraphics[]{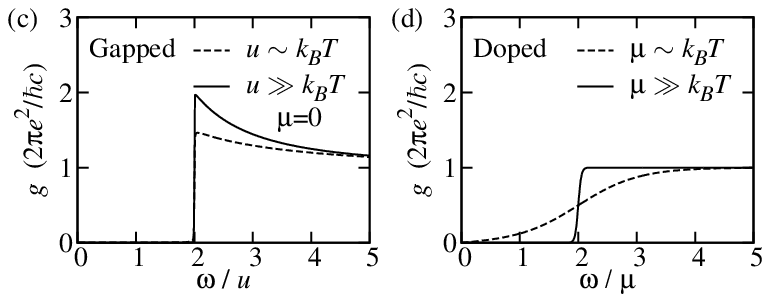}
	\caption{Absorption coefficient in the FIR frequency range for (a)
	undoped strained BLG (or nematic state) with $w=5\mathrm{meV}$,
	$T=3\mathrm{K}$; 
	(b) the doped, strained case with $w=5\mathrm{meV}$,
	$T=3\mathrm{K}$, and $\mu=2\mathrm{meV}$ and $6\mathrm{meV}$.  
	(c) The undoped gapped phase.
	(d) The unperturbed case with $\mu>0$.
	\label{fig:fir}}
\end{figure}

We now turn our attention to absorption in the FIR frequency range,
$|\omega| \sim 2w \ll \gamma_1$, where the relevant optical transitions
occur between the two low-energy bands.
Figure~\ref{fig:fir}(a) illustrates features of the absorption by
strained BLG for the two characteristic polarizations of FIR radiation,
$\ve \parallel \vell$ and $\ve \perp \vell$: a weak polarization
dependence described by $Q\approx +0.3$ at energies $\omega \approx 2w$
indicating that the absorption is strongest for light polarized in the
direction of the principal strain axis. 
Note that for FIR light the relation between absorption in different
polarizations is opposite to what we found for the IR spectral range.
A dip in the absorption at very low energies is due to the finite
temperature. 
The absorption by strained BLG with finite doping is shown in
Fig.~\ref{fig:fir}(b). The only effect of the doping is to cut off the
absorption for $\omega < |\mu|$. The absorption by the gapped phase is
shown in Fig.~\ref{fig:fir}(c) for comparison: it has no polarization
dependence but has a peak at $\omega=2u$ corresponding to the threshold
of the optical transition.
The doped but unperturbed BLG, Fig.~\ref{fig:fir}(d)
shows a step in $g(\omega)$ at $\omega = 2\mu$.\cite{nicol-prb77}
Therefore, one can distinguish the type of symmetry breaking in BLG
using FIR spectroscopy through the weak absorption anisotropy of the
strained/nematic state and the band-edge peak in the gapped state.

In conclusion, we have shown that IR and FIR absorption spectroscopy can
distinguish between the gapless anisotropic nematic (or strain-induced)
and the isotropic gapped broken symmetry states of BLG. 
The former has a characteristic strong dependence on the orientation of
the polarization of the incident radiation, such that in the IR
frequency range, light polarized perpendicularly to the strain axis (or
symmetry-breaking axis in the nematic phase) will be absorbed very
strongly at $\omega \approx \gamma_1+w$ in the undoped system, whereas
light polarized parallel to this axis acquires a characteristic feature
in absorption when the doping is such that $\mu>w$. 
There is also a weak polarization dependence for strained BLG in the FIR
regime, with the parallel polarization being absorbed more strongly. 
In contrast, the isotropic gapped phases show no absorption anisotropy,
but do have qualitatively different lineshape from both strained BLG and
the unperturbed state at zero or finite doping.

This study was supported by US-ONR and NRI-SWAN (DSLA) and by the EPSRC,
the European Research Council, and the Royal Society (VIF). 
It was conducted at the KITP ``Physics of Graphene'' program,
supported in part by the National Science Foundation under grant No.
NSF PHY11-25915.

\bibliography{bibtex-sorted}
\end{document}